\documentclass[manuscript]{acmart}

\usepackage{booktabs} 
\usepackage{subcaption,graphicx}
\usepackage{stackengine}
\stackMath
\usepackage{tikz}
\usepackage{pgf-pie}
\usepackage{pgfplots}
\usepgfplotslibrary{groupplots}
\usepackage{tabularx}
\theoremstyle{definition} 
\newtheorem{defn}{Definition} 

\newcolumntype{P}[1]{>{\centering\arraybackslash}p{#1}}
\setcopyright{none}

\author{Mohammad Akif Beg}

\affiliation{Concordia University}

\author{Jia Yuan Yu}

\affiliation{Concordia University}

\begin{document}
\title{Generating Embroidery Patterns Using Image-to-Image Translation}

\begin{abstract}
In many scenarios in computer vision, machine learning, and computer graphics, there is a requirement to learn the mapping from an image of one domain to an image of another domain, called {\em Image-to-image translation}. For example, {\em style transfer}, {\em object transfiguration}, visually altering the appearance of weather conditions in an image, changing the appearance of a day image into a night image or vice versa, photo enhancement, to name a few. In this paper, we propose two machine learning techniques to solve the embroidery image-to-image translation. Our goal is to generate a preview image which looks similar to an embroidered image, from a user-uploaded image. Our techniques are modifications of two existing techniques, {\em neural style transfer}, and {\em cycle-consistent generative-adversarial network}. Neural style transfer renders the semantic content of an image from one domain in the style of a different image in another domain, whereas a cycle-consistent generative adversarial network learns the mapping from an input image to output image without any paired training data, and also learn a loss function to train this mapping. Furthermore, the techniques we propose are independent of any embroidery attributes, such as elevation of the image, light-source, start, and endpoints of a stitch, type of stitch used, fabric type, etc. Given the user image, our techniques can generate a preview image which looks similar to an embroidered image. We train and test our propose techniques on an embroidery dataset which consist of simple 2D images. To do so, we prepare an unpaired embroidery dataset with more than $8000$ user-uploaded images along with embroidered images. Empirical results show that these techniques successfully generate an approximate preview of an embroidered version of a user image, which can help users in decision making.
\end{abstract}

%
%



\maketitle

\section{Introduction}

Customizable fashion is on the rise in fashion industry. Embroidery customization is considered to be the most popular type of customization in comparison to the counterparts, screenprint and digital print. For this work we partnered with a firm that personalize apparels with custom embroidery. A customer uploads an image to be embroidered. Having a real-time approximate preview of their design is a significant factor in the decision making of the client. Based on our industry research and knowledge, we did not find any work which simulates an approximate embroidered version of a user-uploaded image automatically to facilitate the customization experience. The image-to-image translation is one of the techniques that can help us in providing an approximate embroidered version of an image.
\begin{figure}[h]
  \includegraphics[width=.45\columnwidth]
    {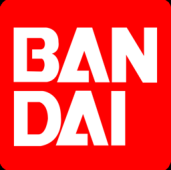}\hfill
  \includegraphics[width=.45\columnwidth]
    {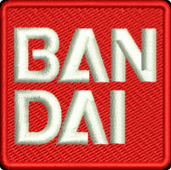}

  \caption{Embroidery image-to-image translation: The image on the left is a an image from source domain, whereas the image on the right is from translated domain.}
  \label{Embroidery}
\end{figure}
\par
The image-to-image translation is a process of translating one possible representation of an image to another, while keeping the original structure and semantics of the image (such as a change in style,  design alteration, colorization, and others) intact. Alterntively stated, the objective is to learn the mapping from an image of one domain to an image of another domain  Before the introduction of this technique, to achieve a similar task a combination of different image processing techniques were required \cite{isola2017image} like image quilting \cite{efros2001image}, image analogies \cite{hertzmann2001image}, image denoising \cite{buades2005non}, depth prediction \cite{eigen2015predicting}, semantic labeling \cite{eigen2015predicting}, surface normal estimation \cite{eigen2015predicting}, image colorization \cite{zhang2016colorful} and many others \cite{shih2013data, laffont2014transient, xie2015holistically, chen2009sketch2photo}. \par In this work, we propose to solve embroidery image-to-image translation, where a 2D image is translated to an embroidered version of the image without altering the high level semantic content. Figure (\ref{Embroidery}), shows an example of an embroidery image-to-image translation. We propose to solve this problem by modification of two existing techniques, neural style transfer \cite{gatys2015neural} and cycle-consistent generative adversarial networks (CycleGAN) \cite{Zhu_2017}. Both of these techniques have shown remarkable results in solving different image-to-image translation problems. Gatys et al. \cite{gatys2015neural} introduced neural style transfer to solve an image-to-image translation problem for art. They used a convolutional neural network (CNN) \cite{simonyan2014very} to transfer the style of a painting on a photograph. Since then, many different researchers have proposed different modifications and advancements to the original algorithm, and all of them have achieved groundbreaking results. Our work is directly related to the line of work initiated by Gatys. Another technique that we propose to solve the embroidery image-to-image translation problem is using cycle-consistent generative adversarial networks. Goodfellow et al. \cite{NIPS2014_5423} introduced generative adversarial networks (GAN). They proposed a framework which uses an adversarial process for estimating the generative model. Different architectures and modification of generative adversarial networks are used to solve image-to-image translation problems. Isola et al. \cite{isola2017image}  as a general-purpose solution to image-to-image translation problem proposed conditional adversarial networks. They have tested this framework on a paired set of images, and the results were quite remarkable. Zhu et al. \cite{Zhu_2017} used an unpaired set of images and cycle-consistent adversarial network to solve the image-to-image translation problem. The use of an unpaired set of images has given this framework an advantage to be widely used in different applications. Though there are many different approaches to solving an image-to-image translation problem, our work is closely related to the work of Zhu et al. \cite{Zhu_2017} because of the freedom of using unpaired images.

\section{Baseline Techniques}\label{baseline}
Image-to-image translation is one of the prominent problem in computer vision and graphics. Many different approaches have been propose by different researchers to solve different applications of image-to-image translation. For the embroidery image-to-image translation, we use to existing machine learning techniques specially design to solve image-to-image translation. We modify these techniques for our specific application of embroidery. 

\begin{figure*}
  \includegraphics[width=\textwidth,height=5cm]{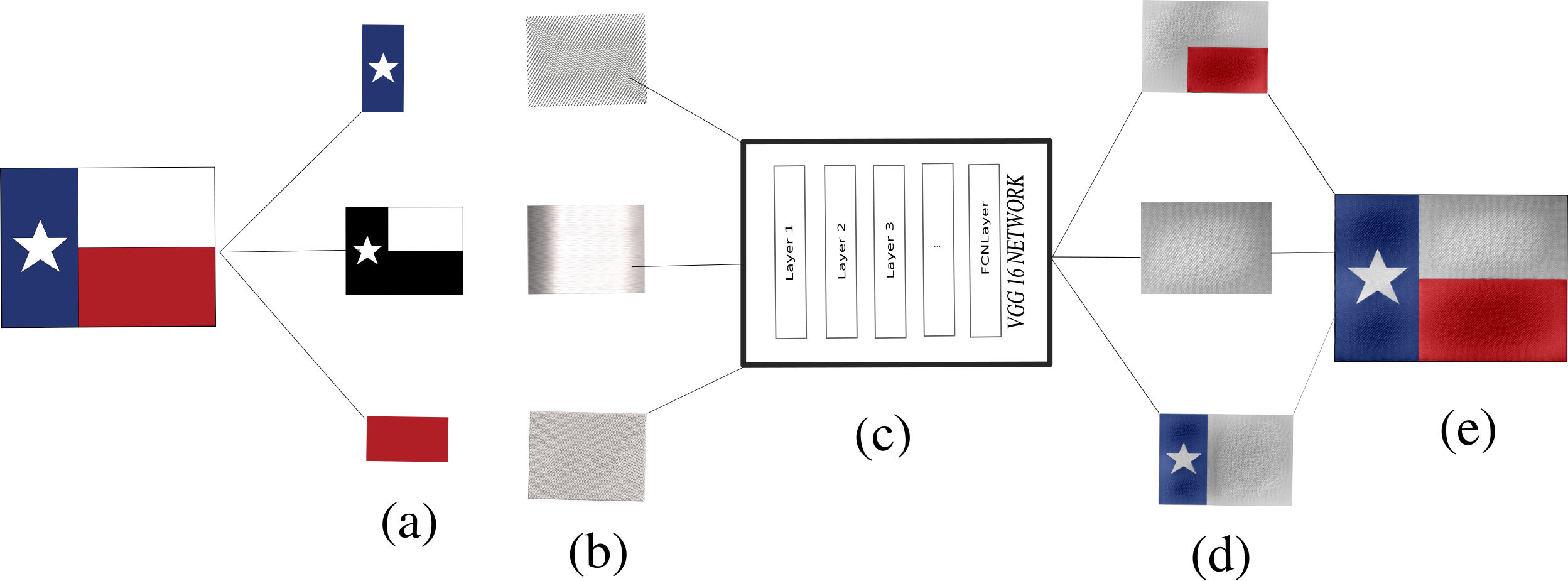}
  \caption{Split Style Transfer: this figure demonstrate the technique we propose to create an embroidered version of an image. The algorithm is separated into 5 steps. (a) split the image into different sub-images of different colors. (b) select different types of stitch for each sub-image. (c) perform style transfer on sub-images using different stitches. (d) styled sub-images using different stitches. (e) combine the styled sub-images. Note: The sub-image in (b) is made with a black background on purpose for understanding.}
  \label{fig:sst}
\end{figure*}

\subsection{Neural Style Transfer}
Neural style transfer is a technique proposed by Gatys et al. \cite{gatys2015neural}, where the objective is to transfer the style of one image onto another image without changing its high-level semantic content. The main approach of neural style transfer is to jointly minimize the distance of the style representation and the content representation, which is learned in different layers of the convolution neural network (CNN). The authors of \cite{gatys2015neural}, advocate that we can distinguish layers that are responsible for representing the style and the ones responsible for representing the content of an image and we can separate these layers to work independently on the content and style and therefore can achieve the image-to-image translation. Neural style transfer defines a loss function that tries to minimise the differences between a content image, a style image and a generated image.
\begin{defn}[Feature Map]\label{def:feature} 
Let $d^{l}$ is the number of filters in the $l^{th}$ layer of the CNN. $\hat{u}^{l}$ and $\hat{v}^{l}$ be the height and weight of the feature map at layer $l$, $p^{l}$ represents the size of the feature map at layer $l$. $p^{l}$ is the product of $\hat{u}^{l} \times \hat{v}^{l}$. Let $z$ be any given image, we define the feature map as:
\begin{equation}
H^{l}_{{ij}}(z) \in \mathbb{R}^{d^{l} \times p^{l}},
\end{equation}where $H^{l}_{{ij}}$ is the activation of the $i^{th}$ filter at position $j$ in layer $l$.
\end{defn}
Feature maps are a good representation of the features of input image. They capture spatial information of an image without containing the style information. The feature maps helps in defining the loss function of neural style transfer.
\begin{defn}[Gram Matrix]\label{def:gram}
Given an image $z$, we define gram matrix as the dot product of the feature map defined above in (\ref{def:feature}).
\begin{align}
G(y)^{l}_{ij} = \sum_{k=1}^{p^{l}}H^{l}_{ik} \cdot H^{l}_{jk}.
\end{align}
\end{defn}
Flattened feature vectors from a convolutional feature map representing features of the input space, and their dot product give us the information about the style or texture of an image and this dot product is referred as gram matrix.

\subsection{Cycle-Consistent Generative Adversarial Network}
The second method we propose is inspired by cycle-consistent adversarial network (CycleGAN) introduced by Zhu et al. \cite{Zhu_2017}, which is an extension of generative adversarial network (GAN) \cite{NIPS2014_5423}. CycleGAN has achieved substantial success in the image-to-image translation problem domain because it uses an unpaired image dataset. However, Isola et al. \cite{isola2017image} have already introduced a conditional adversarial network which provides promising results, but it requires a paired image dataset and creating paired image dataset for a new task like ours is a tedious process. CycleGAN uses a pair of generators to achieve the translation problem of images from input domain X to an output domain Y and also it can translate the images from output domain Y to input domain X. The main concept of CycleGAN is to provide translation from the original domain to the target domain and vice versa. There are two components to the CycleGAN loss function, an adversarial loss and a cycle consistency loss. Both are essential to getting good results. An adversarial loss is similar to GANs, in which the generator is trying to "fool" the corresponding discriminator. The adversarial loss alone is not sufficient to produce good images, as it leaves the model under-constrained. It enforces that the generated output to be of the appropriate domain, but does not enforce that the input and output are recognizably the same. The cycle consistency loss addresses this issue. It relies on the expectation that if you convert an image to the other domain and back again, by successively feeding it through both generators, you should get back something similar to what you put in.
Recently, there has been many new modifications done to the architecture of CycleGAN to improve the quality of generated results, we combined two such modifications to solve embroidery image-to-image translation problem.

\section{Our Approaches}
\subsection{Split Style Transfer}
Split style transfer is a modification to neural style transfer \cite{gatys2015neural} technique, discussed in previous section (\ref{baseline}). During our experiments, neural style transfer was manage to generate some decent results for a single color image but for multi-color images, the results are a bit different from the approximate preview. When an image is actually embroidered on an apparel, every distinct color in an image is embroidered using a different type of stitch pattern or a different orientation of a similar stitch pattern to distinguish surfaces. Generating approximate embroidery version using neural style transfer has the restriction of using only single style image. To achieve more realistic approximate preview, we propose to split a multi-color 2D image into sub-images of distinct colors. Then apply different style transfer on each of the sub-image using a different type of embroidery patterns and combine the subimages at the end. The benefit of split style transfer is that it ensures all distinct colors in an image have different embroidery pattern and the final combined generated image is a more realistic approximate preview. We also made some modifications in the loss function of neural style transfer. Figure (\ref{fig:sst}) shows the split style transfer method we propose to translate embroidery on a given image.

\subsubsection{Problem Formulation} 
Let $u \in Z^{+}$ be the height, $v \in Z^{+}$ be the width and $c \in Z^{+}$ be the color channel of each image. For $r = \{1,\cdots,e\}$, let $x$ be the content image, such that $x \in \mathbb{N}^{u \times v \times c}$ and $Y = \{y_{1}, \cdots, y_{e}\}$ represents the set of style image (\textit{embroidery images in our case}), such that $y_{i} \in \mathbb{N}^{u \times v \times c}$ similar to neural style transfer. In split style transfer, we first split our content image $x$, into a set of images of distinct color $X^{\prime} = \{x^{\prime}_{1}, \cdots, x^{\prime}_{e}\}$, where $e$ is the total number of distinct colors in an image. Then, every sub image $x^{\prime}_{r}$ is styled using a different style image, $y_{r} \in Y$. The final styled images, $\hat{Y}^{\prime} = \{\hat{y}_{1}, \cdots \hat{y}_{e}\}$ are then combined to one image $\hat{y}$ at the end after style transfer is performed separately. We define the content $L^{l}_{c}$, style $L^{l}_{s}$, and total loss $L^{l}_{t}$ of split style transfer for a layer $l$ as :\\
\textbf{Content Loss}
At any layer $l$, content loss is defined as the mean squared difference of the feature map [\ref{def:feature}] of the generated image and the feature map of the original content image.
\begin{equation}
L^{l}_{c}(x^{\prime}_{r} , \hat{y}^{\prime}_{r}) = \sum_{r=1}^{e} \bigg(\frac{1}{d^{l}p^{l}} \sum_{i=1}^{d^{l}} \sum_{j=1}^{p^{l}} (H(\hat{y}^{\prime}_{r})^{l}_{{ij}} - H(x^{\prime}_{r})^{l}_{{ij}})^{2} \bigg)
\end{equation}
\textbf{Style Loss}
At any layer $l$, style loss is defined as the mean squared difference of the feature map [\ref{def:gram}] of the generated image and the feature map of the original content image.
\begin{equation}
L^{l}_{s}(y_{r} , \hat{y}^{\prime}_{r}) = \sum_{r=1}^{e} \bigg(\frac{1}{d^{l}p^{l}} \sum_{i=1}^{d^{l}} \sum_{j=1}^{p^{l}} (G(\hat{y}^{\prime}_{r})^{l}_{{ij}} - G(y_{d})^{l}_{{ij}})^{2} \bigg)
\end{equation}
\textbf{Total Loss}
At any layer $l$, the total loss of split style transfer is the combination of the content loss and the style loss. The hyperparameters $\alpha$ and $\beta$ are the weights of content and style loss, respectively. By controlling $\alpha$ and $\beta$ we can control the amount of content and style present in the styled image.
\begin{equation}
L^{l}_{t}(x^{\prime}_{r}, y_{r} , \hat{y}^{\prime}_{r}) = \alpha L^{l}_{c}(x^{\prime}_{r} , \hat{y}^{\prime}_{r}) + \beta L^{l}_{s}(y_{r} , \hat{y}^{\prime}_{r})
\end{equation}

\begin{figure*}[h]
\includegraphics[width=\textwidth, height = 8cm]{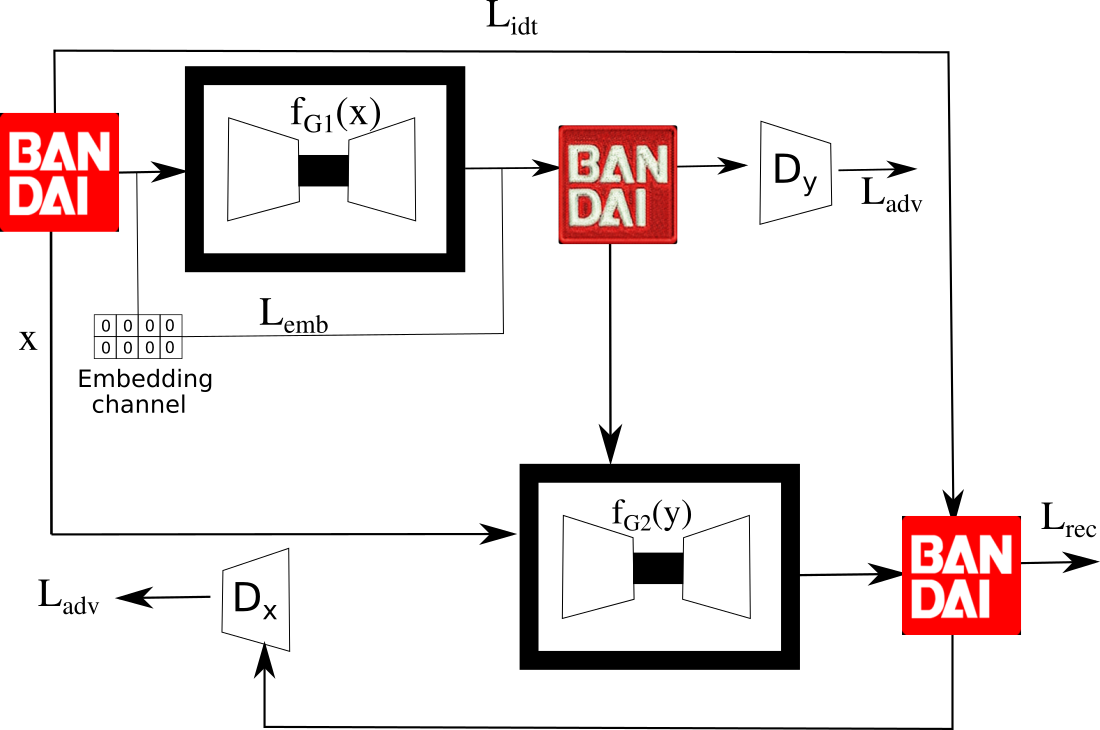}
\caption{A simplified architecture of our EmbGAN. For simplicity, we have just included one cycle from $x$ to $\hat{y}$ to $x^{\prime}$. The eight zero's represents the embedding channel which is added to the input and output of the generator. $L_{adv}$ are the adversarial losses, $L_{emb}$ is the embedding loss, $L_{idt}$ is the identity loss. The complete architecture along with loss functions are explained in section (\ref{modify}), page (\pageref{modify}).}
\label{fig:modigan}
\end{figure*}

\subsection{Embroidery Cycle Consistent GAN (EmbGAN)}
The second technique we propose is inspired by the architecture of CycleGAN \cite{Zhu_2017}. We propose some modifications to the architecture of CycleGAN for embroidery translation problem. Instead of using just instance normalization \cite{ulyanov2016instance} like the original CycleGAN, we have added spectral normalization \cite{miyato2018spectral} similar to \cite{zhang2018pet} to improve the quality of the generated image. We also took inspiration from \cite{longman2018embedded}, which uses a regularized embedded channel for both input and output of the generators. The embedded channel helps the generator memorize the important structures necessary for the reconstruction of the image. As the embedded channel is regularized, it eliminates the possibility of the generator learning the entire image and hence not depreciating the cycle-consistency of CycleGAN. Combining these two modifications to the original architecture of CycleGAN definitely help a lot in improving the quality of the generated images in comparison to original CycleGAN, which we have shown in the following section. Let us mathematically formulate the embroidery translation problem. Each image from the input domain has multiple visual attributes like color, shape, size, luminance, texture, and many others. In the CycleGAN \cite{Zhu_2017} paper, the author has mentioned that there is a primal interconnection between the input image and the generated image. In the embroidery translation problem, the input image and the generated image have many similarities. Ideally, almost everything except the texture of the images is similar.  We will introduce the modified loss function in the following section.

\subsubsection{Problem Formulation} 
Let the dataset consists of user images and embroidered images, both of the same dimension. Let $u \in Z^{+}$ be the height, $v \in Z^{+}$ be the width and $c \in Z^{+}$ be the color channel of each image. Suppose n be the total number of images in the dataset and $N = \{1,2,\cdots,n\}$. Let $X = \{x_{1},x_{2},\cdots,x_{n}\}$ be the set of user images, where $x_{i} \in \mathbb{N}^{u \times v \times c}$ for $i \in N$. Let $Y = \{y_{1},y_{2},\cdots,y_{n}\}$ be the set of embroidered images, where $y_{i} \in \mathbb{N}^{u \times v \times c}$ for $i \in N.$ For $i \in \{1,2\}$, let  $G_{i}$ be a generator and $D_{i}$ be a discriminator, they consist of neural networks with parameters $\theta_{G_{i}}$ and $\theta_{D_{i}},$ respectively, $\theta_{G_{i}}$ and $\theta_{D_{i}}$ are the set of weights and biases of neural networks. Let $Q \times Q$ be the dimension of the noise vector $z_{i} \in Z$. For $i \in \{1,2\}$, we define differentiable function function $f_{G_{i}}:\mathbb{N}^{u \times v \times c} \times \mathbb{R}^{Q \times Q} \rightarrow \mathbb{N}^{u \times v \times c}$. We denote the set of images generated by the generator as $\hat{Y} = \{\hat{y_{1}},\hat{y_{2}},\cdots,\hat{y_{n}}\}$, where $\hat{y}_k \in \mathbb{N}^{u \times v \times c}$ for $k \in N$. Similarly, $f_{D_{i}} : \mathbb{N}^{u \times v \times c} \times \mathbb{N}^{u \times v \times c} \rightarrow [0,1]$ represents discriminator functions. Here, $f_{D_{i}}(\theta_{D_{i}})$ and $f_{G_{i}}(\theta_{G_{i}})$ are neural networks. $f_{D_{i}}(\theta_{D_{i}})$ used the sigmoid activation function, and $f_{G_{i}}{\theta_{G_{i}}}$ uses the hyperbolic tangent function as activation to get the desired outputs. We define the sigmoid function to classify the set of data points in two desired labels. It predicts the probability of occurrence of a particular label say $y \in \{0,1\}$. Let $x$ be the input, $w$ be the weight vector and $b$ be the bias of the neural network. We defined the sigmoid function $s : \mathbb{R} \rightarrow [0,1]$ as
\begin{align*}
s(z) = \dfrac{1}{1 + \exp^{-z}},
\end{align*}
where $z = w \cdot x + b$. If the probability of occurrence of the label is 1 is $s(z)$ then the probability of occurrence of label 0 is $1-s(z)$. Let $x^{'}$ be the input, $w^{'}$ be the weight vector and $b^{'}$ be the bias of another neural network, we define the hyperbolic tangent function $tanh : \mathbb{R} \rightarrow [-1,1]$ as following
\begin{align*}
tanh(u) = \dfrac{e^{u} - e^{-u}}{e^{u} + e^{-u}},
\end{align*}
where $u = w^{'} \cdot x^{'}+b^{'}$ \cite{haykin2004comprehensive}. We use cross entropy to compute the similarity measure between the probability of occurrence of labels based on the true data points from the training dataset and the probability of occurrence of labels from the generated data points. Let  $P =\{P_{1},P_{2},\ldots,P_{n}\}$ be the visual attributes of an image $\in \{x,y\}$ and, $Q =\{Q_{1},Q_{2},\ldots,Q_{n}\}$ be the values of these attributes. Our goal is to find two optimized mappings $f_{G^{*}_{1}} : X \rightarrow Y$ and $f_{G^{*}_{2}} : Y \rightarrow X$. For $\forall x \in X,$ the generated image $\hat{y} = f_{G_{1}}(\theta_{G_{1}},x)$, the values $Q_{x}$ and $Q_{y}$ of all attributes $P$ should be kept same except the embroidery attribute. That means the optimal mapping $f_{G^{*}_{1}}$ transferred only the embroidery attribute without destroying any other attribute \cite{zhang2018pet}.

\textbf{Loss Function}
\label{modify}
The loss function of our model is different from our baseline CycleGAN \cite{Zhu_2017}. To achieve better quality results, we have taken inspirations from different researchers and modified the architecture of CycleGAN to achieve the best results for embroidery translation problem. 
The loss function for our model is as follows: The adversarial loss and the cyclic loss is similar to CycleGAN. We have also introduced identity loss, and embedded loss. We have used a L1 norm loss function for the additional channel to the input image which is defined as the absolute difference between the target value and the estimated value. For $i  = \{1,2,\cdots,n\}$, let $S$ be the difference, $t_{i}$ be the target value and $p_{i}$ be the estimated value, we define L1-norm as:
\begin{align*}
S = \sum_{i=1}^{N} | t_{i} - p_{i} |.
\end{align*}
\textbf{Adversarial Loss:} Each generator in the model tries to minimize its loss, whereas each discriminator in the model tries to maximize its loss. The adversarial loss focuses on the fact that the data distribution between the output domain and generated domain measures up with each other. Let the set of input images be X = $\{x_{1}, x_{2}, \ldots,x_{n}\}$ and the set of output images be Y = $\{y_{1}, y_{2}, \ldots,y_{n}\}$. Suppose $L_{j}(\theta_{G_{j}},\theta_{D_{j}})$ be the adversarial loss for the generative adversarial network $j$, for $j \in \{1,2\}$ and $x_{i} \in X, y_{i} \in Y, z_{i} \in Z$ for all $i \in \{1,2,\ldots,N\}$. We define adversarial loss as follows:
 
\begin{align}
\begin{split}
L_{1}&(\theta_{G_{1}},\theta_{D_{1}})= \frac{1}{N} \sum_{i=1}^{N} \log f_{D_{1}}(\theta_{D_{1}},x_{i},y_{i})P(Y=y_{i})\\
& + \frac{1}{N} \sum_{i=1}^{N} \log(1-f_{D_{1}}(\theta_{D_{1}}, x_{i},f_{G_{1}}(\theta_{G_{1}},x_{i})))P(X=x_{i}), \label{eqn:adversarialloss1}
\end{split}
\end{align} 
for all $\theta_{D_{1}}$ and $\theta_{G_{1}}$, and
\begin{align}
\begin{split}
L_{2}&(\theta_{G_{2}},\theta_{D_{2}}) = \frac{1}{N} \sum_{i=1}^{N} \log f_{D_{2}}(\theta_{D_{2}}, y_{i},x_{i})P(X=x_{i})\\
& + \frac{1}{N} \sum_{i=1}^{N} \log(1-f_{D_{2}}(\theta_{D_{2}}, y_{i},f_{G_{2}}(\theta_{G_{2}},y_{i})))P(Y=y_{i}), \label{eqn:adversarialloss2}
\end{split}
\end{align}
for all $\theta_{D_{2}}$ and $\theta_{G_{2}}$. The goal here is to find the optimal value of $\theta^{*}_{G_{j}}$ and $\theta^{*}_{D_{j}}$, $\forall j$.
We can state the objective as follows:
\begin{align}
\begin{split}
\theta^{*}_{G_{j}},\theta^{*}_{D_{j}} = \arg\stackunder{\min}{\theta_{G_{j}}}  \stackunder{\max}{\theta_{D_{j}}} L_{j}(\theta_{G_{j}},\theta_{D_{j}}), \forall j.
\end{split}
\end{align}
\textbf{Cycle Consistency Loss:} CycleGAN introduces a cyclic approach to converting the generated image back to its subsequent image from the original domain. The loss incurred during the process is addressed as Cycle Consistency Loss. Since we have two generators and discriminator pairs, we also have two cyclic consistency loss, namely:
\begin{enumerate}
\item Forward Consistency Loss: An image $x_{i} \in X$ domain is fed to the generator $f_{G_{1}}(\theta_{G_{1}}x_{i},y_{i})$ which generates $y^{'}_{i}$ and then this is again fed to generator $f_{G_{2}}(\theta_{G_{2}},y^{'}_{i},x_{i})$ which generates $x^{'}_{i}$, ideally close to $x_{i} \forall i$. Forward consistency loss is defined as:
\begin{align}
\begin{split}
&L_{cyc_{1}}(\theta_{G_{1}}, \theta_{G_{2}}) =\\& \frac{1}{N} \sum_{i=1}^{N} |f_{G_{2}}(\theta_{G_{2}}, f_{G_{1}}(\theta_{G_{1}}, x_{i},y_{i}),x_{i})-x_{i} |P(X=x_{i}) \label{eqn:fwd}
\end{split}
\end{align}
\item Backward Consistency Loss : Similarly to forward consistency loss, we define backward consistency loss as follows:
\begin{align}
\begin{split}
&L_{cyc_{2}}(\theta_{G_{2}}, \theta_{G_{1}}) =\\& \frac{1}{N} \sum_{i=1}^{N} |f_{G_{1}}(\theta_{G_{1}}, f_{G_{2}}(\theta_{G_{2}}, y_{i},x_{i}),y_{i})-y_{i} |P(Y=y_{i}) \label{eqn:bckwrd}
\end{split}
\end{align}
\end{enumerate} 
\textbf{Identity Loss}: Identity loss is an optional loss in the original CycleGAN paper, but during our training process, we learned that adding an identity loss helps in training the network efficiently similar to \cite{zhang2018pet}. Identity loss helps the generator identify whether the image is from the input domain or the output domain and hence reducing the chances of making it just a mapping algorithm. We define identity loss as:
\begin{align}
\begin{split}
L_{idt}(\theta_{G_{1}},\theta_{G_{2}})= \frac{1}{N} \sum_{i=1}^{N} f_{G_{2}}((\theta_{G_{2}}, x_{i})-x_{i}) + f_{G_1}((\theta_{G_{2}},{y_{i}})-y_{i}). \label{eqn:iden}
\end{split}
\end{align}
\textbf{Embedding Loss}: We have added an embedded channel to the input and output of both the generators to assist in the process of learning the structure of the input image as shown in the figure (\ref{fig:modigan}),  is able to learn a separate channel, which can be thought of as an embedding, that helps in the reconstruction of the input. The idea is to encourage the generator to generate a properly translated image and an embedding of the changes that were made are required to reconstruct the input. An L1 distance is used with the embedding loss, similar to \cite{Zhu_2017}. The reason of using an L1 distance instead of L2 is that L1 produces less blurring images \cite{isola2017image}. Also, since a L2-norm squares the error (increasing by a lot if error $>$ 1), the model will see a much larger error ( e vs $e^2$ ) than the L1-norm, so the model is much more sensitive to this example, and adjusts the model to minimize this error. If this example is an outlier, the model will be adjusted to minimize this single outlier case, at the expense of many other common examples, since the errors of these common examples are small compared to that single outlier case. Let the extra embedded channel to the input of the image is donated by H, we define the embedding loss as 
\begin{align}
\begin{split}
L_{emb}(\theta_{G_{1}},\theta_{G_{2}})= \frac{1}{N} \sum_{i=1}^{N} ||f_{G_1}(\theta_{G_{1}},(x_{i}+H))||_{1}+||f_{G_2}(\theta_{G_{2}}, y_{i}+H)||_{1}. \label{eqn:emb}
\end{split}
\end{align}
Our goal is to obtain optimal parameters for the sum of all losses. Our final loss is defined as:
\begin{align}
\begin{split}
&\theta^{*}_{G_{j}},\theta^{*}_{D_{j}} = arg\stackunder{min}{\theta_{G_{j}}}  \stackunder{max}{\theta_{D_{j}}} \sum_{j=1}^{2}(L_{j}(\theta_{G_{j}},\theta_{D_{j}}) + L_{cyc_{j}}(\theta_{G_{j}},\theta_{G_{k}})\\ 
&+\lambda_{1}L_{idt}(\theta_{G_{j}}) + \lambda_{2} L_{emb}(\theta_{G_{j}}),\\
&\forall j, \forall k \in \{1,2\} \wedge j \neq k. \label{eqn:final}
\end{split}
\end{align}

\section{Experiment}
\subsection{Dataset}
\begin{figure}[h]
\begin{tabular} {ccc}
{\includegraphics[width = 1.0in, height= 3cm]{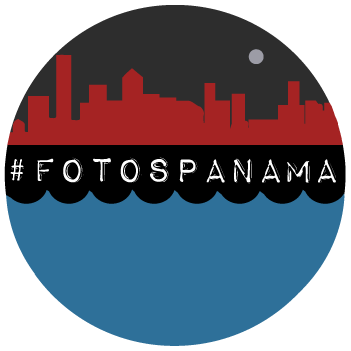}} & 
{\includegraphics[width = 1.0in, height= 3cm]{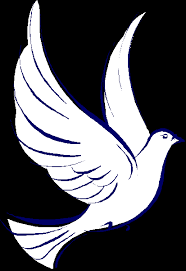}} &
{\includegraphics[width = 1.0in, height= 3cm]{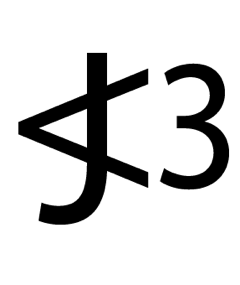}} \\
{\includegraphics[width = 1.0in, height= 3cm]{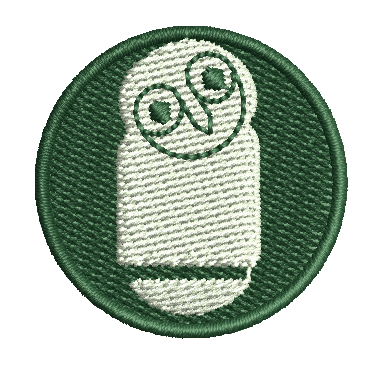}} &
{\includegraphics[width = 1.0in, height= 3cm]{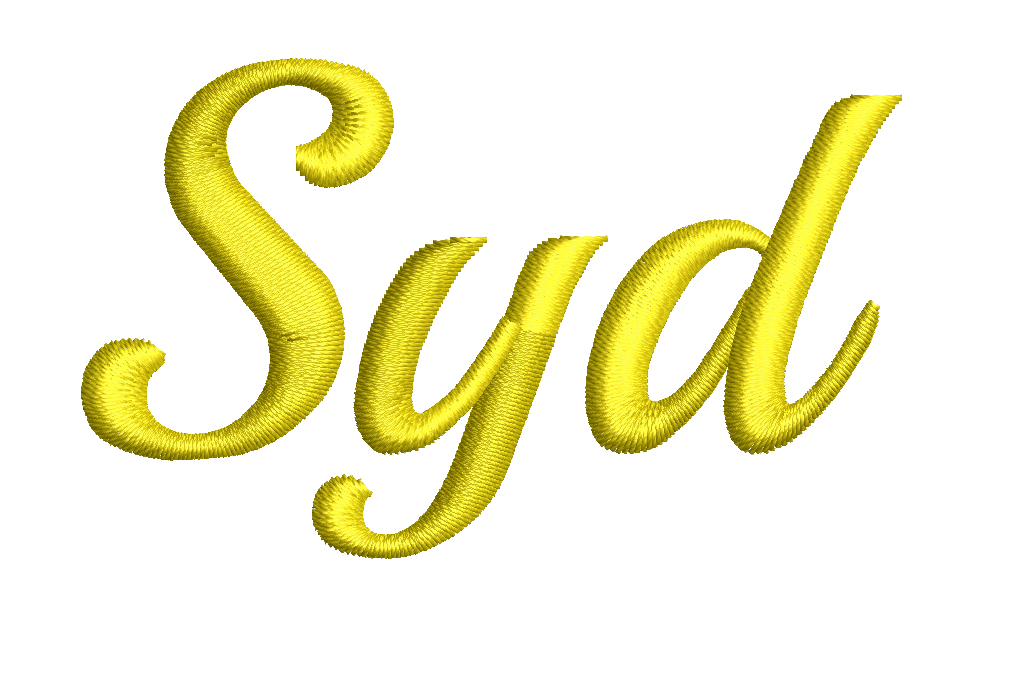}} &
{\includegraphics[width = 1.0in, height= 3cm]{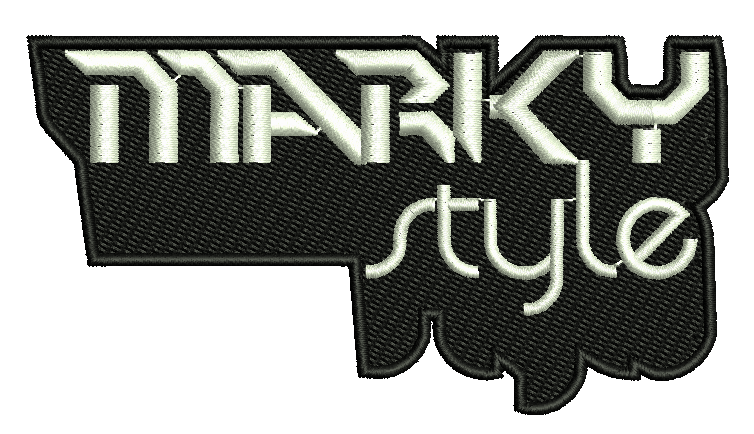}} \\
\end{tabular}
\caption{Sample images from our dataset. 1st row :  User uploaded images, 2nd row : manually embroidered images using proprietary software.}
\label{images}
\end{figure}

We have prepared a dataset of 2D images for embroidery image-to-image translation. The image used in the dataset are simple two dimensional images which can be embroidered have fairly less complex semantic content. A photographic image of a skyline or an image with complex semantic content would be feasible for embroidery and hence is not used in preparing the dataset. Most of the images are simple flags, texts, logs or other simple structural images as shown in \ref{images}. We can broadly divide the dataset into two categories, one of which is the textual image which consists of word(s) that a user has uploaded to be customized on their apparel and the other category are any non-textual images. The other division of the dataset is based on the color of the images, we have multi-color images, single-color images and gray-scale images. The total number of images in the dataset is 8668, out of which 4643 are user uploaded images, and 4025 are manually embroidered version of these images.
\begin{table}[h]
\centering
\caption{Data distribution in embroidery dataset.}
\small
\begin{tabular}{|p{2cm}|p{2cm}|p{1cm}|p{1cm}|p{1cm}|}
\hline
\textbf{Category} & \textbf{Total} & \textbf{Train} & \textbf{Test}\\[5pt]
\hline
\textbf{Textual Images} &   2713  & 1764   & 949\\[5pt]
\hline
\textbf{Non-Textual Images}  & 5955 & 3929 & 2026\\[5pt]
\hline
\textbf{overall}  &8668 &5693  & 2975\\[5pt]
\hline
\end{tabular}
\end{table}

\begin{table}[h]
\caption{AMT "Real vs. Fake" Test. We compare the images generated from CycleGAN and EmbGAN with the ground truth. A turker receives a pair of images where one image is from ground truth, and one is a generated image from either CycleGAN or EmbGAN. The table displays how many times an algorithm was able to fool the turkers.}
\begin{center}
\begin{tabular}{|l||P{60mm}|}\hline \hline
Algorithm &  \% Turkers marked as real \\ \cline{1-2}
CycleGAN & $ 4.08 $\\ \cline{1-2}
EmbGAN & $  14 $ \\ \hline\hline
\end{tabular}
\label{Tab:turker2}
\end{center}
\end{table}

\subsection{Perceptual Studies}
The absolute best metric for evaluating the results of any image-to-image translation problem or any other image synthesization task in the field of computer graphics and computer vision is to check how compelling the results are to a human observer. Similar to the baseline architecture, we have run perceptual studies of "real vs. fake" on Amazon Mechanical Turk, the protocol used is identical to [\cite{isola2017image}, \cite{Zhu_2017}, \cite{zhang2016colorful}]. We have also run a comparison perceptual studies comparing the results of baseline architecture and our modified architecture. For the "real vs. fake" perceptual studies, the turkers were presented with a series of pair of images, where one of the images is ground truth which is considered "real," and the other image is generated by our algorithm which is considered as "fake" in our study. The pair of images appeared for 1 second, and after the images disappeared, the turkers were given an unlimited amount of time to decide which one was real or fake. The Table \ref{Tab:turker2}, displays the result of the first perceptual study. We can see that the images generated from EmbGAN were manage to fool the turkers 4 times more than the images generated from CycleGAN. For the comparative perceptual study, we have increased the time to 2 seconds, so that users will have more time to make the comparison. The Table \ref{Tab:turker3}, displays the result of comparison test between CycleGAN and EmbGAN. 91.5\% of the turkers choose the images generated from EmbGAN to be more realistic as an embroidered image in comparison to the image generated from CycleGAN. Approximately 50 turkers evaluated each algorithm. For every algorithm, we provided a pair of 50 images, but for the comparative study of neural style transfer and split style transfer, we have used a pair of 40 images. The Table \ref{Tab:turker1}, displays the result of the comparison test between neural style transfer and split style transfer. 67.5\% of the turkers choose the images styled by split style transfer to be more realistic as an embroidered image in comparison to the image styled by neural style transfer. To ensure that the participants were competent at this task, we have used the strategy similar to \cite{zhang2016colorful}, 10\% of the trials pitted the ground truth with the images generated from initial epochs and the turkers were able to identify the generated image as fake 94\% of the time, indicating that the turkers understand the task in hand. To ensure that there is no bias and all the algorithms were tested in the similar environment, we ensure that all the experiments were carried out at the same time of the day and all the sessions were independent and identically distributed to the turkers simultaneously.

\begin{table}[h]
\caption{AMT Comparison Test. We compare the images styled by neural style transfer and split style transfer to one another. A turker receives a pair of images where one image is a styled image by neural style transfer, and the other image is a styled image by split style transfer. The table displays which results were more realistic as an embroidered image, according to the turkers.}
\begin{center}
\begin{tabular}{|l||P{50mm}|}\hline \hline
Algorithm &  \% Turkers marked as real \\ \cline{1-2}
Neural Style Transfer & $ 32.5 $\\ \cline{1-2}
Split Style Transfer & $ 67.5 $ \\ \hline\hline
\end{tabular}
\label{Tab:turker1}
\end{center}
\end{table}

\begin{figure*}[t]
\centering
    \begin{subfigure}[b]{0.13\textwidth}            
            \includegraphics[width=\textwidth,keepaspectratio=true]{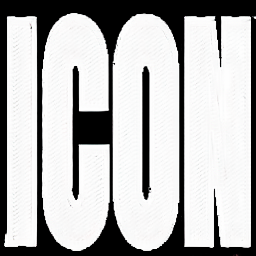}
    \end{subfigure}
    \hspace{0.05mm}
    \begin{subfigure}[b]{0.13\textwidth}            
            \includegraphics[width=\textwidth,keepaspectratio=true]{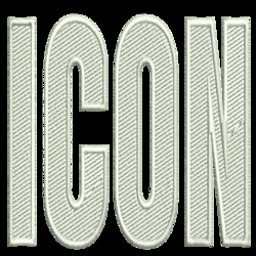}
    \end{subfigure}
    \hspace{0.05mm}
    \begin{subfigure}[b]{0.13\textwidth}            
            \includegraphics[width=\textwidth,keepaspectratio=true]{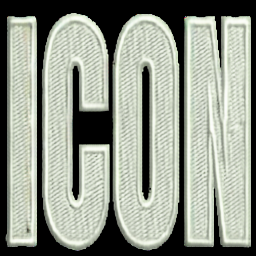}
    \end{subfigure}
    \hspace{0.05mm}
    \begin{subfigure}[b]{0.13\textwidth}            
            \includegraphics[width=\textwidth,keepaspectratio=true]{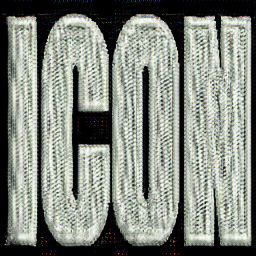}
    \end{subfigure}
    \hspace{0.05mm}
    \\
     \vspace{1mm}
     \begin{subfigure}[b]{0.13\textwidth}            
            \includegraphics[width=\textwidth,keepaspectratio=true]{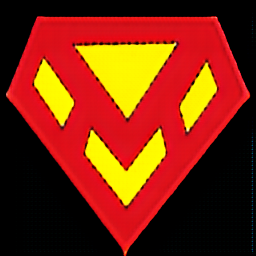}
    \end{subfigure}
    \hspace{0.05mm}
    \begin{subfigure}[b]{0.13\textwidth}            
            \includegraphics[width=\textwidth,keepaspectratio=true]{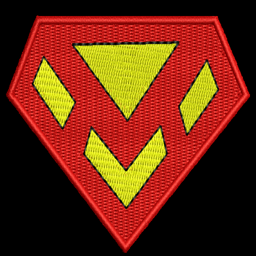}
    \end{subfigure}
    \hspace{0.05mm}
    \begin{subfigure}[b]{0.13\textwidth}            
            \includegraphics[width=\textwidth,keepaspectratio=true]{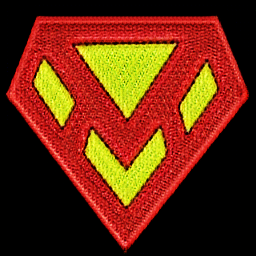}
    \end{subfigure}
    \hspace{0.05mm}
    \begin{subfigure}[b]{0.13\textwidth}            
            \includegraphics[width=\textwidth,keepaspectratio=true]{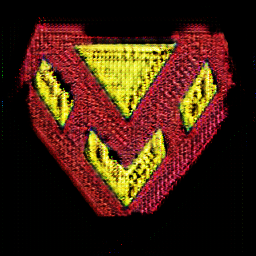}
    \end{subfigure}
    \hspace{0.05mm}
    \\
     \vspace{1mm}
     \begin{subfigure}[b]{0.13\textwidth}            
            \includegraphics[width=\textwidth,keepaspectratio=true]{lateximages/input1.png}
    \end{subfigure}
    \hspace{0.05mm}
    \begin{subfigure}[b]{0.13\textwidth}            
            \includegraphics[width=\textwidth,keepaspectratio=true]{lateximages/groundtruth1.png}
    \end{subfigure}
    \hspace{0.05mm}
    \begin{subfigure}[b]{0.13\textwidth}            
            \includegraphics[width=\textwidth,keepaspectratio=true]{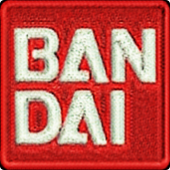}
    \end{subfigure}
    \hspace{0.05mm}
    \begin{subfigure}[b]{0.13\textwidth}            
            \includegraphics[width=\textwidth,keepaspectratio=true]{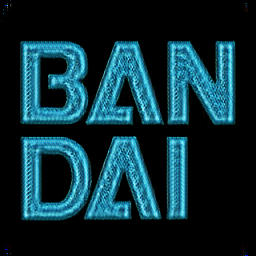}
    \end{subfigure}
    \hspace{0.05mm}
    \\
     \vspace{1mm}
     \begin{subfigure}[b]{0.13\textwidth}            
            \includegraphics[width=\textwidth,keepaspectratio=true]{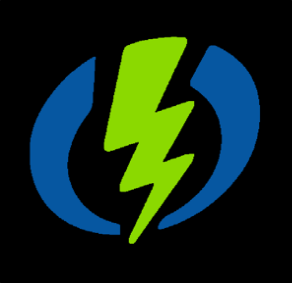}
    \end{subfigure}
    \hspace{0.05mm}
    \begin{subfigure}[b]{0.13\textwidth}            
            \includegraphics[width=\textwidth,keepaspectratio=true]{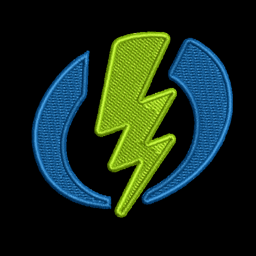}
    \end{subfigure}
    \hspace{0.05mm}
    \begin{subfigure}[b]{0.13\textwidth}            
            \includegraphics[width=\textwidth,keepaspectratio=true]{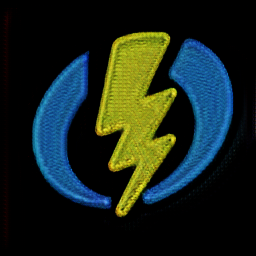}
    \end{subfigure}
    \hspace{0.05mm}
    \begin{subfigure}[b]{0.13\textwidth}            
            \includegraphics[width=\textwidth,keepaspectratio=true]{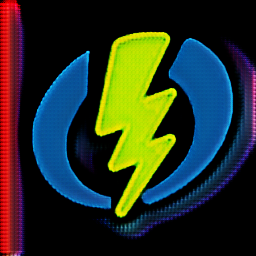}
    \end{subfigure}
    \hspace{0.05mm}
    \\
     \vspace{1mm}
     \begin{subfigure}[b]{0.13\textwidth}            
            \includegraphics[width=\textwidth,keepaspectratio=true]{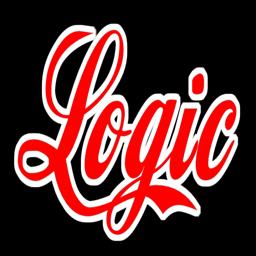}
            \caption*{Input Images}
    \end{subfigure}
    \hspace{0.05mm}
    \begin{subfigure}[b]{0.13\textwidth}            
            \includegraphics[width=\textwidth,keepaspectratio=true]{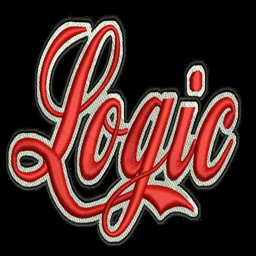}
            \caption*{Ground Truth}
    \end{subfigure}
    \hspace{0.05mm}
    \begin{subfigure}[b]{0.13\textwidth}            
            \includegraphics[width=\textwidth,keepaspectratio=true]{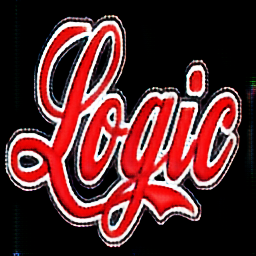}
            \caption*{Our Method}
    \end{subfigure}
    \hspace{0.05mm}
    \begin{subfigure}[b]{0.13\textwidth}            
            \includegraphics[width=\textwidth,keepaspectratio=true]{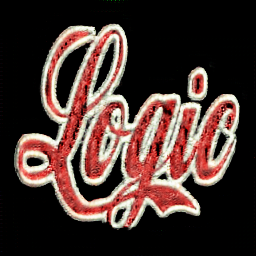}
            \caption*{CyleGAN}
    \end{subfigure}
    \hspace{0.05mm}
    \\
     \vspace{1mm}
    \hspace{0.05mm}
    \caption{EmbGAN Result: Every subfigure has four versions. One of the them is the user-uploaded input image. Second, is the manually digitized embroidered version of the image used as ground truth. Third, is the result generated by original CylceGAN. Lastly, the result generated by EmbGAN. Note: The results are best to observe in color.}
    \label{figure:gan}
\end{figure*}

\begin{figure}[h]
  \centering
  \begin{subfigure}{.3\columnwidth}
    \centering
    \includegraphics[width=\linewidth]{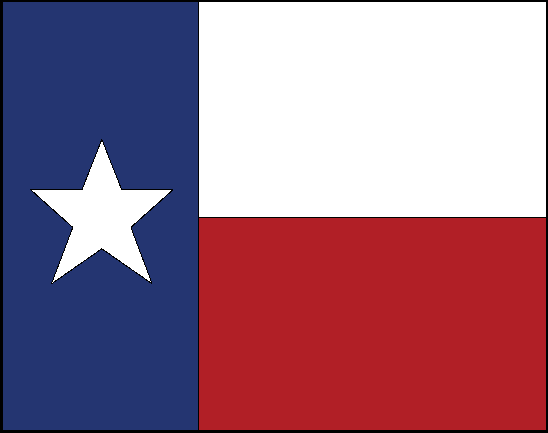}
  \end{subfigure}%
  \hfill
  \begin{subfigure}{.3\columnwidth}
    \centering
    \includegraphics[width=\linewidth]{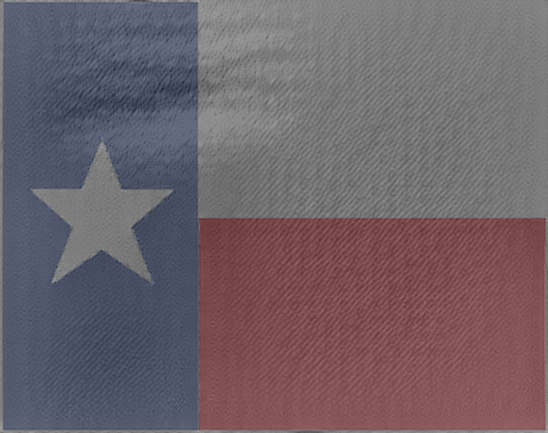}
  \end{subfigure}%
  \hfill
  \begin{subfigure}{.3\columnwidth}
    \centering
    \includegraphics[width=\linewidth]{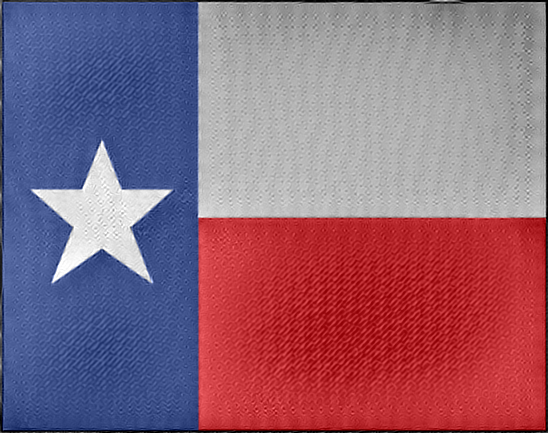}
  \end{subfigure}
  \\
  \vspace{2mm}
  \begin{subfigure}{.3\columnwidth}
    \centering
    \includegraphics[width=\linewidth]{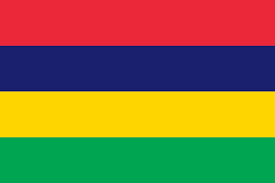}
  \end{subfigure}%
  \hfill
  \begin{subfigure}{.3\columnwidth}
    \centering
    \includegraphics[width=\linewidth]{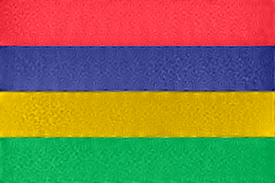}
  \end{subfigure}%
  \hfill
  \begin{subfigure}{.3\columnwidth}
    \centering
    \includegraphics[width=\linewidth]{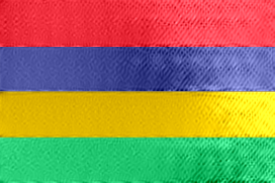}
  \end{subfigure}
  \\
  \vspace{2mm}
  \begin{subfigure}{.3\columnwidth}
    \centering
    \includegraphics[width=\linewidth]{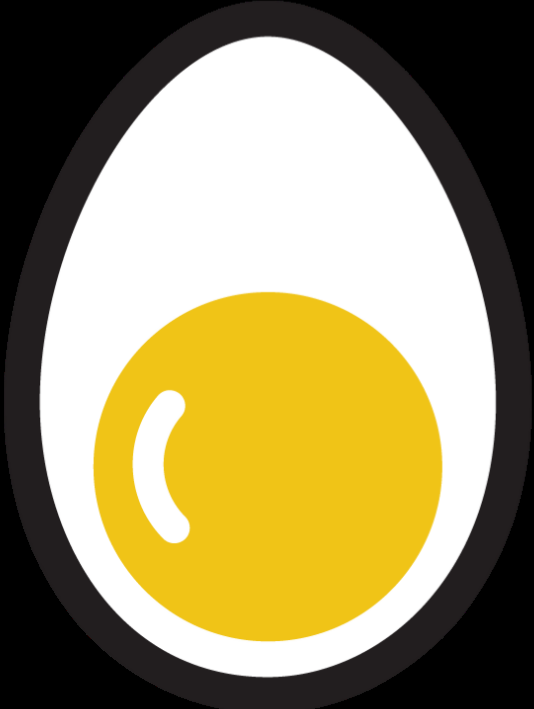}
    \caption{User}
  \end{subfigure}%
  \hfill
  \begin{subfigure}{.3\columnwidth}
    \centering
    \includegraphics[width=\linewidth]{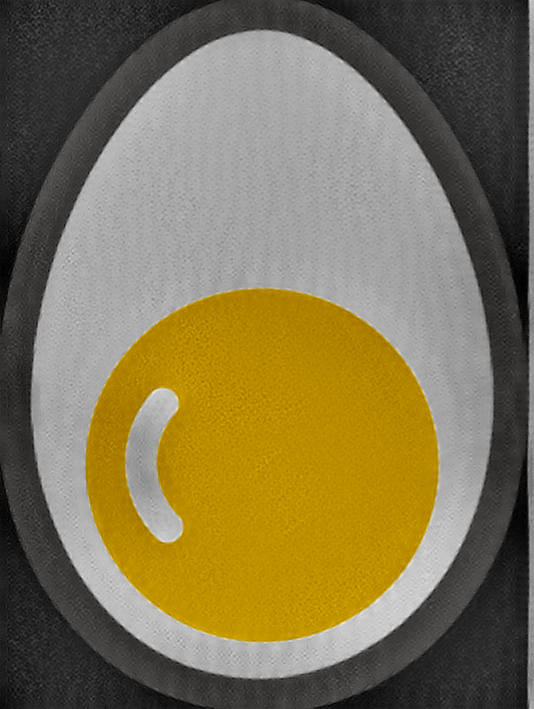}
    \caption{Neural}
  \end{subfigure}%
  \hfill
  \begin{subfigure}{.3\columnwidth}
    \centering
    \includegraphics[width=\linewidth]{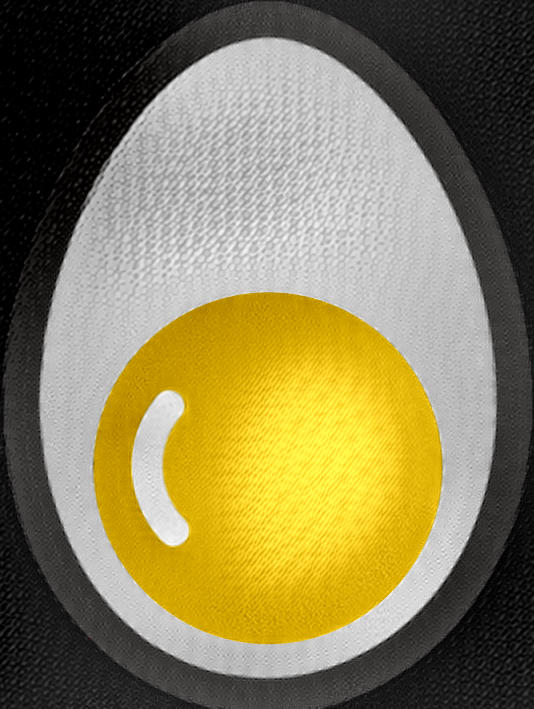}
    \caption{Split}
  \end{subfigure}
  \caption{Style Transfer : (a) is the user uploaded image, (b) is a styled image using neural style transfer and one embroidery image, (c) is a style image using split style transfer and three different embroidery image for each distinct color. Note: The results are best to observe in color.}
  \label{figure:sst}
\end{figure}

\subsection{Split Style Transfer Results}
Figure \ref{figure:sst} shows the images styled by neural style transfer and our propose method split style transfer. Every image has three versions, the first is the user uploaded image, the second is a styled image using neural style transfer and one embroidery image, the third is a style image using split style transfer and three different embroidery image for each distinct color. The results will help us in perceptually compare the algorithms based on the visual quality of the styled images.

\begin{table}[h]
\caption{AMT Comparison Test. We compare the images generated from CycleGAN and EmbGAN to one another. A turker receives a pair of images where one image is a generated image from CycleGAN, and the other image is a generated image from EmbGAN. The table displays which results were more realistic as an embroidered image, according to the turkers.}
\begin{center}
\begin{tabular}{|l||P{60mm}|}\hline \hline
Algorithm &  \% Turkers marked as real \\ \cline{1-2}
CycleGAN & $ 8.5 $\\ \cline{1-2}
EmbGAN & $ 91.5 $ \\ \hline\hline
\end{tabular}
\label{Tab:turker3}
\end{center}
\end{table}

\subsection{EmbGAN Results}
Figure \ref{figure:gan} shows the images generated by EmbGAN. We have compare them with the images generated by the original CycleGAN. Also, we will compare both these generated images with the images digitized by proprietary software which is considered to be the ground truth and see how our well both methods perform in generating an approximate embroidered version of an image. Each subfigure has four versions. One of them is the user-uploaded input image. Second, is the manually digitized embroidered version of the image used as ground truth. Third, is the result generated by original CycleGAN. Lastly, the result generated by EmbGAN.

\section{Conclusion}
In this work, we propose two techniques to solve our embroidery image-to-image translation problem. The techniques we propose are a modification of two existing machine learning techniques which are popular in producing good results for any given image-to-image translation task,  \textit{neural style transfer} and \textit{cycle-consistent generative adversarial network}. We have done an embroidery image-to-image translation to generate an approximate real-time preview for a customer who wants to have customized embroidery on their apparel. The approximate preview of the final embroidered version of their uploaded two-dimensional image will help the customer in the decision-making process and hence will help in reducing the amount of product returned because of customer's dissatisfaction.  The results from both the techniques were very satisfactory. We used perceptual studies to compare the results with the baseline architectures and the results from the split style transfer were qualitatively better than neural style transfer, and the generated images from EmbGAN were qualitatively better than CycleGAN. We also compared the generated images from both the techniques which conclude that the EmbGAN generates the best quality embroidered image for any given image. We have used the digitized embroidered version of images using industry software as the baseline for comparison. After our final comparison between the generated images, we find out that, the style transfer technique is not the best approach for solving embroidery image-to-image translation. The images generated by style transfer feel less realistic visually, and hence, would not be the best approximate preview of the embroidered version of a user-uploaded image.

\bibliographystyle{ACM-Reference-Format}
\bibliography{bibfile} 

\end{document}